\documentclass[12pt,preprint]{aastex}

\newcommand{\msol}{\mathrm{M_{\odot}}}
\newcommand{\thco}{^{13}\mathrm{CO}}
\newcommand{\twco}{^{12}\mathrm{CO}}
\newcommand{\kms}{{\mathrm{km \, s^{-1}}}}

\newcommand{\de}{{^{\circ}}}
\newcommand{\gs}{\mathrel{\raise0.35ex\hbox{$\scriptstyle >$}\kern-0.6em
\lower0.40ex\hbox{{$\scriptstyle \sim$}}}}
\newcommand{\ls}{\mathrel{\raise0.35ex\hbox{$\scriptstyle <$}\kern-0.6em
\lower0.40ex\hbox{{$\scriptstyle \sim$}}}}

\slugcomment{Submitted to Ap J L }

\shorttitle{Starbursts in Milky Way}
\shortauthors{Stark et al.}

\begin{document}

\title{Gas Density, Stability and Starbursts Near the\\ 
Inner Lindblad Resonance of the Milky Way}

\author{Antony A. Stark, Christopher L. Martin, Wilfred M. Walsh, Kecheng Xiao, Adair P. Lane}
\affil{Harvard-Smithsonian Center for Astrophysics, 60 Garden St., Cambridge MA 02138}
\email{aas@cfa.harvard.edu}

\and

\author{Christopher K. Walker} 
\affil{Steward Observatory, University of Arizona, Tucson, AZ 85721}
\email{cwalker@as.arizona.edu}

\begin{abstract}
A key project of the Antarctic Submillimeter Telescope and
Remote Observatory reported by \citet{martin04} is the mapping
of CO $J=4\rightarrow3$ and $J=7\rightarrow6$ \, emission from the inner 
Milky Way, allowing determination of gas density and temperature.
Galactic center gas that \citet{binney91}
identify as being on $x_2$ orbits has a density 
near $10^{3.5} \, \mathrm{cm ^{-3}}$,
which renders it only marginally stable
against gravitational coagulation into a few Giant Molecular Clouds,
as discussed by \citet{elmegreen94}.
This suggests a relaxation oscillator mechanism for starbursts
in the Milky Way, where inflowing gas accumulates in a ring at 150 pc radius 
for approximately 20 million years, until the critical density is reached, 
and the resulting instability leads to the sudden formation of giant clouds
and the deposition of $4 \times 10^7 {\mathrm{M_{\sun}}}$ of gas onto 
the Galactic center.
\end{abstract}

\keywords{Galaxy: structure---ISM: clouds---ISM: molecules---galaxies: starburst---stars:formation}

\newpage

\section{Introduction}

Dynamics of gas in the inner few kiloparsecs of the
Milky Way are dominated by the non-axisymmetric gravitational 
potential of the central bar.   The properties of this
bar are now well known, and there is good agreement
between observations of 
gas motion and model fits to the potential 
\citep{jenkins94,gerhard99,hafner00,bissantz03}. 
As suggested by \citet{binney91}, the gas tends to
be found on families of closed orbits which are 
not self-intersecting.  All non-closed orbits 
and some closed orbits
are self-intersecting.
Gas on such orbits will shock and lose energy
where the gas streamlines intersect, and the
gas will then move inwards to a lower energy orbit.
If the gas can find its way onto a family of 
non-self-intersecting closed orbits, 
the energy dissipation slows and the timescale for
orbital changes lengthens out.
\citet{contopoulos77} described
two families of closed orbits in barred galaxies:
the ``$x_1$'', which are elongated along the bar
and found outside the inner
Lindblad resonance (ILR); 
and the ``$x_2$'', which are more round
and can be found near the ILR and inside it.
The ILR is located where the epicyclic frequency of
a particle orbiting in the Galactic potential
resonates with the pattern speed of the bar.
This occurs at a radius of
approximately 450 pc from the center of the Milky Way
\citep{bissantz03}.  

Gas which is several kiloparsecs 
away from the Galactic center tends to be driven inwards
until it reaches a region within a few hundred
parsecs of the ILR, because the interaction of
the bar potential with the gas exerts a
negative torque, resulting in loss of angular momentum
by the gas
\citep{lyndenbell72,athanassoula88,jenkins94}. 
Near the ILR this effect disappears, because the net torque there is 
small or zero,
and inwards of the
ILR it may even reverse and become positive, so that
gas inside the ILR could be driven outwards \citep{combes88}.
Gas therefore accumulates in a ring near the ILR.
Unlike the gas further out, the dynamics of this gas 
depends critically on its self gravity \citep{elmegreen94,jenkins94},
and therefore on its thermodynamic properties, 
density in particular.

The thermodynamic properties of the gas can be determined by 
millimeter- and sub\-mil\-li\-meter-wave spectral line observations.
The distribution of
molecular gas near the ILR is known from extensive
surveys in CO and $\thco$ $J=1\rightarrow0$ and
$J=2\rightarrow1$ \citep{bally88a,bitran97,oka98};  these
spectral lines show the presence of molecular gas.
These lines alone do not, however, determine its
density or excitation temperature.
Observations of the mid-$J$ lines of CO 
provide the missing information. 
Since the low-$J$ states of CO are in local thermodynamic
equilibrium (LTE) in almost all molecular gas \citep{goldreich74},
measurements of mid-$J$ states are critical to
achieving a solution
of the radiative transfer
by breaking the degeneracy between beam filling factor
and excitation temperature.
A new survey \citep{martin04}
by the Antarctic Submillimeter Telescope and Remote Observatory 
\citep[AST/RO,][]{stark01} adds
the $J=4\rightarrow3$ (461 GHz) and $J=7\rightarrow6$ (807 GHz) 
rotational lines 
of CO to existing lower-frequency data \citep{bally88a}.
These data are available on 
the AST/RO website\footnote{\tt http://cfa-www.harvard.edu/ASTRO} 
for general use.
These measurements have recently been modeled
using the large velocity gradient
(LVG) approximation to determine the gas density and temperature.

In this {\em Letter}, we discuss the implications of the \citet{martin04}
density estimates for Galactic center gas. 
We apply our new data, specific to the Milky Way, to the
general analysis of stability of dense gas near ILR regions
in galaxies by \citet{elmegreen94}.
We find that the gas near 150 pc radius is marginally unstable.
This suggests that in the past there has been a period of stability 
and gas build-up.
In the future, the instability will create a few giant clouds,
resulting in a starburst and the deposition of tens of millions
of solar masses of material on the Galactic center.
This process repeats with a cycle time determined by the rate at
which gas precipitates on the Galactic center region from outside,
resulting in 
starbursts at intervals of approximately 20 million years.

\section{Gas Density}

Figure 1 is a representation of the average density of the molecular
gas layer in the vicinity of the ILR.
\citet{martin04} used an LVG model on their survey 
data to estimate the density and
temperature at each point in ($\ell$, $b$, $v$).
Density values in Figure 1 are calculated by 
averaging the logarithm of the density 
values from \citet{martin04} at each value of $(\ell, v)$, as $b$
varies from $-0.3$ to $0.2$, excluding points where the LVG fit is
uncertain.   The excluded points are those where the average $\twco$
$J = 1 \rightarrow 0$ emission in \citet{bally88a} is less than 0.2K, or
where the average density is below
$10^{2.5} \, {\mathrm{cm^{-3}}}$, which is 
an approximate lower limit to the
validity of the \citet{martin04} LVG model.
The pixels at some values of $(\ell, v)$ in Figure 1 appear
white, and this can be for several reasons:
(1) there are no data at that position
since the survey is limited in $v$ by the bandwidth 
of the spectrometer;
(2) there is no significant CO emission;
(3) the density is below the threshold of validity of the model;
(4) the variance in the average over $b$ exceeds $(10^3 \, \mathrm{cm^{-3}})^2$.
The model has converged to a consistent 
value at the points where a density is displayed.
Even those values could be spurious if the
assumptions used in the LVG approximation do not apply.
This certainly occurs
where there is foreground absorption far from the 
emitting region, for example 
between $-60$ and $+25 \, \kms$, where spiral arms in the
Galactic disk are projected onto the Galactic center emission.
This region is hatched in Figure 1.

Superposed on the data are some closed orbits from the
model of \citet{bissantz03}.  
The cusped $x_1$ orbit and two self-intersecting $x_1$ orbits
are shown in magenta.  The cusped $x_1$ orbit is the innermost
$x_1$ orbit which does not self-intersect, with an apogee
1300 pc from the Galactic center.  The two self-intersecting
$x_1$ orbits are interior to this, with apogees near 1000 pc.
The $x_2$ orbits begin inside the ILR, and are drawn in
brown.  The outermost $x_2$
has an apogee of 180 pc.  The $x_2$ orbits extend all the way
into the Galactic center; the innermost shown here is almost
circular and has an apogee of 23 pc.
These same orbits are plotted differently in
Figures 10 and 11 of \citet{bissantz03}.

We see that in some places along the cusped $x_1$
orbit, $n({\mathrm{H_2}}) \approx 10^3$ to 
$10^4 \, \mathrm{cm^{-3}}$, although the gas is clumpy
and the average density on this orbit is less. 
Most places along the outer $x_2$ orbit have
$n({\mathrm{H_2}}) \approx 2 \times 10^3$ to 
$6 \times 10^3 \, \mathrm{cm^{-3}}$, 
with few values outside this range.
The densities on the $x_2$ orbits which are at negative $\ell$
and negative $v$ are unreliable, because these parts of the
orbit lie about 1 kpc behind some of the 
outer $x_1$ orbits (not shown in Figure 1)
which have the same velocity but lower excitation.
Most of the region between the cusped $x_1$ orbit 
and the outermost $x_2$ orbit
has lower density and little molecular emission.

\section{Gas Stability}

Given a measure of the gas density on orbits near the ILR, it
is possible to determine whether or not that gas 
is stable---if the gas is 
sufficiently dense, its self-gravity will overcome
the tidal shear of the Galaxy's gravitational potential and it
will agglomerate into clouds.
The dynamics of this situation has been analyzed by
\citet{elmegreen94}, who linearized the hydromagnetic force
and continuity equations in a rotating
frame and thereby derived a dispersion relation for the growth of
instabilities in a gas ring near the ILR. 
He determined that the growth rate
is a function of 
gas density and pressure, the Galactic potential, magnetic field strength, 
and the rate at which material accretes from larger radii.
This relation is expressed in equation 11 of \citet{elmegreen94}: 
\begin{equation}
\omega_r^2 - \omega_G^2 + \Omega_a \omega_r + {{\kappa^2 \omega_r^2}\over{
\omega_r^2 + k^2 v_A^2 +\Omega_a \omega_r}} \approx 0
 \, ,
\end{equation}
where $\omega_r$ is the growth rate of instability, $k$
is its wavenumber, $\kappa$ is the epicyclic frequency in the
Galactic potential, $v_A$ is the Alfv\'{e}n velocity corresponding
to the azimuthal magnetic field, 
$\Omega_a$ is the relative accretion rate, and $\omega_G$
is a frequency related to the acceleration of self-gravity.
Values of the quantities in Equation 1 are estimated in Table 1.

The relative accretion rate, $\Omega_a$,
is a measure of gas inflow from larger radii.  It is determined by
processes outside the ILR region: the amount of gas in the outer
regions of the bar, and the torques exerted on that gas
by the bar.  At minimum, the evolved stars in the bulge will
eject matter into the interstellar medium of the outer bar at
a rate $\sim 0.2 \, \msol \mathrm{yr^{-1}}$
\citep{jungwiert01}.  At maximum,
the Galactic center region could ingest an entire gas-rich dwarf
companion at a rate of 
$100$ to $1000 \, \msol \mathrm{yr^{-1}}$.  Accretion rates
higher than this would disrupt the inner Galaxy, so we can
conclude they have not occurred in 
the Milky Way \citep{heller94,bournaud02}.
\citet{combes04} suggests 
an accretion rate of $10 \, \msol \mathrm{yr^{-1}}$, 
which is approximately the rate of accretion of
intergalactic material onto the Galaxy.  
This rate, accreting onto
inner gas disk of $\sim 2 \times 10^8 \, \msol$, yields
$\Omega_a \sim 50 \, \mathrm{Gyr^{-1}}$.

There is as yet no measurement of the strength of
the magnetic field in the dense Galactic center 
gas, although
the existence of a magnetic field perpendicular to
the disk near the Galactic center is demonstrated 
by the non-thermal radio filaments \citep{yusefzadeh84},
and the existence of a magnetic field in the plane of
the disk is demonstrated by submillimeter-wave polarimetry \citep{novak03}.
\citet{chuss03} have estimated the field strength
to be about a milliGauss, based on the field morphology.
The perpendicular field near the center, they argue, is the
result of a process where the field in the disk
is amplified 
until the magnetic field pressure begins to dominate
the dynamics and the field decouples from the gas
in the regions of relative lower density.
This implies that the magnetic field energy in
the regions of relative higher density self-regulates to
approximate equipartition with the internal kinetic energy of the gas,
making 
the Alfv\`{e}n velocity, $v_A$,
comparable to the internal turbulent velocity of the
molecular gas.
Accepting this argument leads to a value of 
$ v_A $ in Equation 1 which is significant but not dominant. 
In other barred galaxies, the observed field strength averaged over
the central few kiloparsecs 
is typically $10 \, \mu\mathrm{G}$ \citep{beck02}.  Since the
dense gas has a filling factor of $\sim 10^{-2}$, this is 
consistent with a milliGauss field in the dense gas.

The epicyclic frequency, $\kappa$, is more certain.  It depends only on
the rotation curve and its derivative, which are known
within $\pm20\%$.  In Table 1 we adopt values
taken from Figure 1 of \citet{bissantz03}.

\citet{elmegreen91} approximates the equation of state
for the molecular gas as
$\Delta P \approx \gamma_{\mathrm{eff}} c^2 \Delta \rho$,
where the complexities of heating and cooling of the
molecular gas are subsumed in an effective ratio
of specific heats
$\gamma_{\mathrm{eff}} \sim 0.3$ to $2$, and
an effective sound speed $c \sim 10$ to $40\, \kms$.
As pointed out in \citet{elmegreen94}, the
quantity 
$\gamma_{\mathrm{eff}} c^2$ for the actual gas under
analysis can be estimated by considering the
equilibrium between pressure and self-gravity perpendicular
to the plane: 
$h\, \approx \, (2/\mathrm{e}) \gamma_{\mathrm{eff}}^{1/2} c \kappa^{-1} $,
where the scaleheight, $h$, is determined from observations.
Applied to the values in Table 1, this procedure gives
$\gamma_{\mathrm{eff}} \gs 1$ if $c \sim v_A \sim 25 \, \kms$.

The gravitational frequency, $\omega_G$, as
defined by \citet{elmegreen94} is a measure of the
relative importance of self-gravity compared to pressure in the gas over
a scalelength given by wavenumber $k$:
\begin{equation}
\omega_G \approx 2\,\mathrm{G}\mu k^2 \, \mathrm{ln} \left({{2}\over{kh}}\right)\,-\,
k^2\,\gamma_{\mathrm{eff}}c^2 \,
\end{equation}
where $\mu = \rho \pi h^2$ is the mass per unit length in the ring.
The $\omega_G$ has a maximum when
$k_{\mathrm{max}}\approx 2 h^{-1} \mathrm{exp}[-0.5(1+\gamma_\mathrm{eff}c^2/\mathrm{G}\mu)]$,
and the corresponding gravitational frequency,
$\omega_G = \mathrm{G}\mu k_\mathrm{max}^2$ corresponds to the fastest
growing unstable mode.

We are now in a position to evaluate $\omega_r$, the
growth rate of the fastest unstable mode, by
substituting $k_\mathrm{max}$ for $k$.
Equation 1 is quartic in $\omega_r$ with one positive
real solution.
If $\Omega_a \ll \kappa$, and $k_\mathrm{max}v_A \ls \kappa$, as we
see is the case from Table 1, then 
\begin{eqnarray}
\omega_r & \ll &\omega_G \quad \mathrm{if} \quad 
\omega_G \ll \kappa, \mathrm{and} \nonumber\\
\omega_r & \approx & \omega_G \quad \mathrm{if} \quad 
\omega_G \gs \kappa \, . \nonumber
\end{eqnarray}
In the Milky Way at the present time, the accretion term 
$\Omega_a$ is not dynamically significant to the
instability (although it does determine the time
{\em between} instabilities), and the magnetic field
term $v_A$ has only marginal significance, since its
contribution to the dynamics is in approximate
equipartition with the other terms.
The growth rate of the instability is small as long as
$\omega_G \ll \kappa$, but becomes significant when
$\omega_G \sim \kappa$.  The dominant dynamical effect is
the competition between the tidal shear of the Galaxy
and the tendency for the gas ring to clump
with wavelength $\lambda_\mathrm{max}=2\pi/k_\mathrm{max}$.
The criterion for significant instability, $\omega_G > \kappa$,
can be expressed as a threshold in gas density:
\begin{equation}
n(\mathrm{H_2})> n_\mathrm{crit}(\mathrm{H_2}) \approx {{0.2 \kappa^2}
\over{\mathrm{Gm_H}}}
\approx 10^{3.5} \left[{{\kappa}\over{1000 \,\mathrm{Gyr}^{-1}}}\right]^2 \, ,
\end{equation}
which is essentially a density criterion for the formation of molecular
clouds in the presence of Galactic tidal forces \citep{stark89d}.
The region that begins to contract will initially have a mass approximately
equal to $\mu_\mathrm{max}\lambda_\mathrm{max} = 4 \times 10^7 \, \msol$.
As the instability proceeds, the dynamics will become non-linear, and 
move beyond the valid regime of Equation 1.  This merits further study.

\section{Starbursts in the Milky Way}

We see from Figure 1 that gas on the inner $x_1$ orbits exceeds the critical
density threshold in places, and that the gas is clumped.
The gas on the outer $x_2$ orbits is more smooth, but is at the threshold of 
significant instability.  The timescale for clumping to begin is short: 
$1/\omega_r \sim 1 \, \mathrm{Myr}\,.$
Since $\lambda_\mathrm{max} \approx r$, we expect the instability to result in
only  a few large clouds, containing many millions of solar masses.  
The Giant Molecular Cloud
surrounding Sgr B2, which can be seen as a relatively minor density 
enhancement in Figure 1 at $\ell = 0.65\de$, $v = 60 \, \kms$, is such a cloud. 
Because of their large mass, these clouds are subject to dynamical friction 
with the background of stars in the Galactic bulge, and will spiral into 
the center within 1 Gyr \citep{stark91a}.
Dynamical friction can overcome the forces tending to maintain the gas 
in the ILR region, and allow the gas to continue inwards toward the
galactic center, but only if the gas is organized into sufficiently
large self-gravitating clouds.
This suggests a relaxation oscillator mechanism for quasi-periodic
starbursts in the center of the Milky Way.  At first, the Galactic
center region is relatively clear of gas, like the central region
of M31 is now.  Gas precipitates into the region of the bar, either as
mass loss from evolved stars or as the result of cannibalism of smaller
galaxies.  The bar dynamics drive this gas toward the ILR, where it
will tend to accumulate in a ring as long as
$n(\mathrm{H_2}) \ll n_\mathrm{crit}$.
The ring becomes more and more dense as gas continues to precipitate
from larger radii, and eventually the threshold is reached.  A few
giant clouds will form on a relatively short timescale, creating
a starburst, and the giant clouds will be swept toward the center by
dynamical friction, restoring a condition of relatively low density.
The cycle will repeat on timescales of $\Omega_a^{-1} \sim \, 20 \,
\mathrm{Myr}$, but of course this timescale is highly variable and
can be dramatically shortened by events which precipitate a large 
amount of gas.

\acknowledgments
Support was provided by NSF grant OPP-0126090.
We thank our AST/RO colleagues Richard Chamberlin, Jacob Kooi, and
Gregory Wright for extensive help with the instrumentation.

\bibliographystyle{apj}

\clearpage

\begin{deluxetable}{llrrcl}
\tabletypesize{\scriptsize}
\rotate
\tablecaption{Quantities relating to the stability of gas near the ILR. 
\label{tbl-1}}
\tablewidth{0pt}
\tablehead{
\colhead{Quantity} & & \colhead{Value at $r = 150$ pc} & 
\colhead{Value at $r = 450$ pc} & \colhead{Accuracy}
&\colhead{Reference} 
}
\startdata
Max. molecular number density 
     & $n_{\mathrm{max}} (\mathrm{H_2})$ 
     & $6 \, \times \, 10^3\, \mathrm{cm^{-3}}$ 
     & $10^4 \, \mathrm{cm^{-3}}$ 
     & A
     & \citet{martin04}\\
Max. gas density 
     & $\rho_{\mathrm{max}} = 2.4 \cdot \mathrm{m_H} \cdot 
              n_{\mathrm{max}} (\mathrm{H_2}) $ 
     & $178 \, \msol \mathrm{pc^{-3}}$ 
     & $296 \, \msol \mathrm{pc^{-3}}$ 
     & A\\
Relative accretion rate 
     & $\Omega_a$ 
     & $50 \, \mathrm{Gyr^{-1}}$ 
     & $50 \, \mathrm{Gyr^{-1}}$ 
     & C     
     & \citet{combes04}\\ 
Azimuthal magnetic field 
     & $B_\theta$ 
     &  1 mG 
     &  1 mG 
     & C
     & \citet{chuss03} \\
Az. Alfv\`{e}n velocity
     & $ v_A = B_\theta (4\pi\rho_{\mathrm{max}})^{-1/2}$
     & $26 \, \kms$ 
     & $20 \, \kms$ 
     & C\\
Galactic angular velocity 
     & $\Omega $ 
     & $620 \, \mathrm{Gyr^{-1}} $
     & $314 \, \mathrm{Gyr^{-1}} $
     & A 
     & \citet{bissantz03}\\
Derivative of $\Omega $
     & $\mathrm{d}\Omega /\mathrm{d}r$ 
     & $-2500 \, \mathrm{Gyr^{-1} \, kpc^{-1}} $
     & $-490 \, \mathrm{Gyr^{-1} \, kpc^{-1}} $
     & A
     & \citet{bissantz03}\\
Epicyclic frequency
     & $\kappa = (4\Omega^2 + 2 r \Omega \, \mathrm{d}\Omega /
                \mathrm{d}r)^{1/2}$ 
     & $1036 \, \mathrm{Gyr^{-1}} $
     & $506 \, \mathrm{Gyr^{-1}} $
     & A\\
Molecular half-scaleheight
     & $h$ 
     & 21 pc 
     & 56 pc 
     & A
     & \citet{bally88a} \\
Ratio of specific heats 
     & $\gamma_{\mathrm{eff}} c^2 \equiv \Delta P / \Delta \rho 
                \approx (h \kappa {\mathrm{e}})^2 /4 $
     & $874 \, \mathrm{km^2 \, s^{-2}}$   
     & $1483 \, \mathrm{km^2 \, s^{-2}}$ 
     & B  
     & \citet{elmegreen91}\\
Max. mass per length of ring 
     & $ \mu_{\mathrm{max}} =  \pi \rho_{\mathrm{max}} h^2$ 
     & $ 2.5 \, \times \, 10^{8} \mathrm{\msol \, kpc^{-1}}$ 
     & $ 2.8 \, \times \, 10^{9} \mathrm{\msol \, kpc^{-1}}$ 
     & B
     & \citet{elmegreen94}\\
Wavenumber of max. growth 
     & $k_{\mathrm{max}} \approx 2 h^{-1} {\mathrm{exp}}
              [-0.5(1 + \gamma_{\mathrm{eff}} c^2/G\mu_{\mathrm{max}})]$
     & $38 \, {\mathrm{kpc^{-1}}}$
     & $20 \, {\mathrm{kpc^{-1}}}$ 
     & B
     & \citet{elmegreen94}\\
Wavelength of max. growth 
     & $\lambda_{\mathrm{max}} = 2 \pi/k_{\mathrm{max}}$
     & $165 \, {\mathrm{pc}}$
     & $314 \, {\mathrm{pc}}$
     & B
     & \citet{elmegreen94}\\
Gravitational frequency at $k_{\mathrm{max}}$ 
     & $\omega_G = k_{\mathrm{max}}(G \mu_{\mathrm{max}})^{1/2}$ 
     & $1270 \, \mathrm{Gyr^{-1}} $
     & $2286 \, \mathrm{Gyr^{-1}} $
     & B
     & \citet{elmegreen94}\\
Instability growth rate 
     & $\omega_r$
     & $1013 \, \mathrm{Gyr^{-1}} $
     & $2206 \, \mathrm{Gyr^{-1}} $
     & B \\
\enddata

\tablecomments{
The Sun's distance to the Galactic center is taken to be
$R_{\sun} = 8 \, \mathrm{kpc}$. 
The effective sound speed in the interstellar gas is $c$. 
Galactocentric radius is $r$.
The column labeled ``Accuracy" indicates the approximate
errors --- A: $\pm20\%$; B: $\pm50\%$; C: order-of-magnitude. 
}

\end{deluxetable}

\clearpage

\clearpage

\begin{figure}[t!]
\plotone{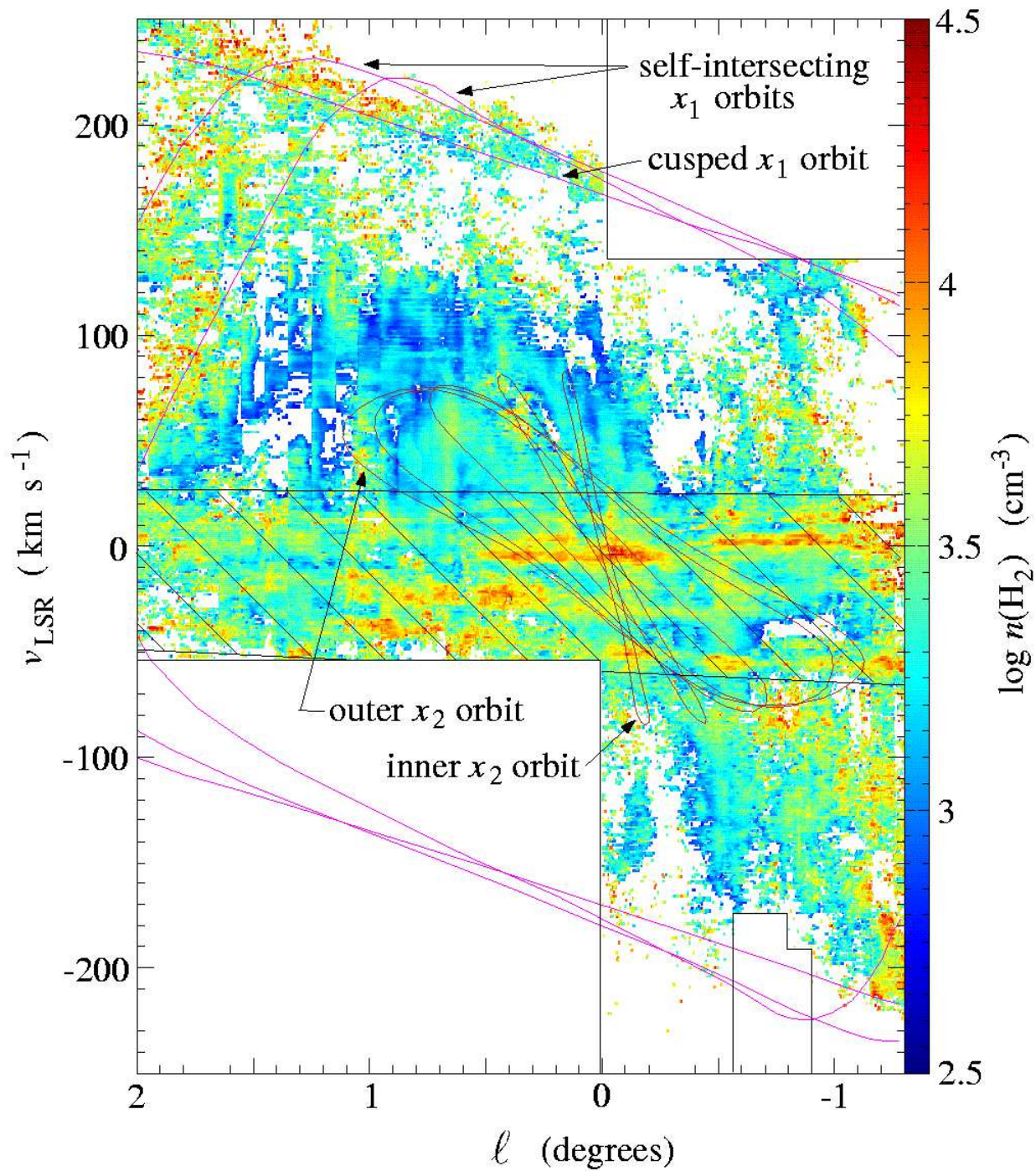}
\caption{Density on $x_1$ and $x_2$ orbits in the inner Milky Way.
The rainbow scale at right shows the average density of molecular gas
from an LVG model using AST/RO survey data \citep{martin04}.
White pixels are points where there are no data, or where the LVG model
does not converge to consistent values.
The hatched area at low velocity 
shows definite foreground absorption which invalidates the assumptions of
the LVG model.  Superposed on the density data are 
some $x_1$ (magenta) and $x_2$
(brown) orbits from \citet{bissantz03}.}
\end{figure}

\clearpage

\end{document}